\documentclass[aps,prl,twocolumn,showpacs]{revtex4}
\usepackage[T1]{fontenc}
\usepackage[latin1]{inputenc}
\usepackage{graphics}
\usepackage{amsmath}



\begin{document}
\title{Synchronized bursts following instability of synchronous spiking
in chaotic neuronal networks}
\author{ Mikhail~V.~Ivanchenko $^{1,2}$, Grigory~V.~Osipov$^{1,2}$,
Vladimir~D.~Shalfeev $^{1}$ and J\"urgen~Kurths$^{2}$}

\affiliation{$^1$Department of Radiophysics, Nizhny Novgorod
University, 23, Gagarin Avenue, 603950 Nizhny Novgorod, Russia,\\
$^2$Institute of Physics, University Potsdam, 10, Am Neuen Palais,
D-14415, Potsdam, Germany }


\begin{abstract}
We report on the origin of synchronized bursting dynamics in
various networks of neural spiking oscillators, when a certain
threshold in coupling strength is exceeded. These ensembles
synchronize at relatively low coupling strength and lose
synchronization at stronger coupling via spatio-temporal
intermittency. The latter transition triggers multiple-timescale
dynamics, which results in synchronized bursting with a
fractal-like spatio-temporal pattern of spiking. Implementation of
an appropriate technique of separating oscillations on different
time-scales allows for quantitative analysis of this phenomenon.
We show, that this phenomenon is generic for various network
topologies from regular to small-world and scale-free ones and for
different types of coupling.
\end{abstract}

\maketitle

Achievements of the recent decade have given strong evidence that
synchronous activity \cite{Pik_book} plays an important role in
the functioning of the nervous system and brain
\cite{neur_review}. These examples range from coordinating
movements in the motor system to information processing
(recognition and perceptual binding) in the visual cortex and
olfactory system. On the other hand, synchronization may play a
destructive role, causing neural disorders like epileptic seizures
or Parkinson's disease \cite{disorders}. Moreover, the importance
of desynchronization in cognitive processing is increasingly being
recognized \cite{desync}.

The need for the theoretical explanation of these findings
stimulated extensive research in the field of nonlinear dynamics.
The most simple and liable to analysis models of
integrate-and-fire neurons, which mimic
periodic subthreshold approach to the spiking state, were
extensively studied to elucidate perfect synchronization
between identical units \cite{Strogatz} and frequency synchronization
between non-identical ones \cite{nonidentical_IF}.
The recently developed theoretical framework for
chaotic synchronization paved the
way to analyzing cooperative dynamics of more realistic models of
chaotically spiking and bursting neurons \cite{models}. Complete
chaotic synchronization in small and large ensembles of {\it
identical} neurons was found and methods of its prediction
were developed \cite{complete}.

However, in nature neurons are {\it not identical}. Therefore, the
functional interdependence between momentary states of
synchronized neurons, if any, becomes extremely complex and
difficult to identify, especially in large ensembles
\cite{Pik_book}. At the same time, as chaotic synchronization has
been observed in a variety of small groups of non-identical
neurons \cite{nonidentical}, it gives strong grounds to expect it
to appear in large ensembles too.

 A promising way to make an advance
here is the concept of {\it chaotic phase synchronization} (CPS)
\cite{CPS}. It implies the adjustment of {\it characteristic
time-scales} of nonidentical oscillators in course of interaction.
Given an appropriate definition of phase and frequency one obtains
an efficacious tool for detecting this process. Neurons, known as
multiple-time-scale systems, can generate either single spikes
mediated by long intervals of silence, or trains of spikes, coined
bursts. Remarkably, as we have recently shown, it is possible to
identify phase synchronization on the bursting time-scale, while
oscillations on the spiking time-scale are unsynchronized
\cite{we_prl}.

In this Letter we study the pathway to the formation of
synchronized bursting in networks of intrinsically spiking
neurons. We show that it is observed with increase of
interneuronal coupling, as the networks achieve synchronous
spiking, undergo its instability towards generation of bursting,
which finally synchronizes. We analyze this phenomenon basing on
the CPS concept and develop a proper technique to separate
oscillations on spiking and burtsting time-scales. We demonstrate,
that, when the CPS regime gets unstable, spatio-temporal
intermittency excites oscillations on the fast time-scale (FTS),
and eventually leads to the regime of synchronized bursts with a
fractal-like spatio-temporal structure of the spikes. Simulations
of scale-free, random, and small-world topologies give evidence of
ubiquity of this phenomenon in complex neuronal networks.

\begin{figure}[t]
{\centering
\resizebox*{0.90\columnwidth}{!}{\includegraphics{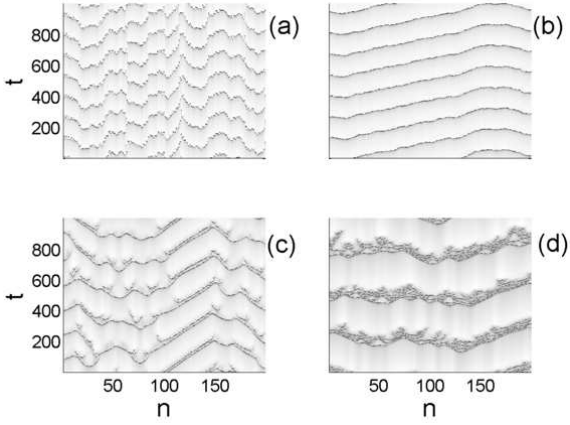}}}
{\caption{Space-time plots illustrate different regimes that occur
in the chain (\ref{eq1}). $x_j$ values are represented by grey
scale, white corresponds to minimal values, black to maximal ones.
Shown are (a) unsynchronized spiking, $\varepsilon=0.01$, (b)
synchronized spiking, $\varepsilon=0.05$, (c) desynchronized
state: synchronization is occasionally broken by fast repetitive
spikes, $\varepsilon=0.1$, (d) synchronized bursts with a
fractal-like spatio-temporal structure of spikes,
$\varepsilon=0.2$.}\label{fig1}}
\end{figure}

We consider networks of non-identical neuronal
model maps, proposed in \cite{Rulkov}:
\begin{equation}
 \label{eq1}
 \left\{
\begin{array}{l}
x_j^{k+1} = f(x_j^k,x_j^{k-1},y_j^k)+ \varepsilon
\sum\limits_{i}G_{j,i}^k/K_i,
 \\
y_j^{k+1} = y_j^k +\mu(- x_j^{k}-1 + \varsigma_j
+\varepsilon\sum\limits_{i}G_{j,i}^k/K_i),
\end{array}
\right.
\end{equation}
\begin{equation}
\label{eq2}
 f(x,\tilde{x},y)=
 \left\{
\begin{array}{l}
\alpha/(1-x) + y, \mbox{ if } x \le 0,\\
\alpha+ y, \mbox{ if } 0 < x <\alpha +y  \ \mbox{and} \\
\ \ \ \ \ \ \ \ \tilde{x} \le 0, \\
 -1, \mbox{ if } x \ge \alpha +
y \ \ \mbox{or} \ \tilde{x} > 0,
\end{array} \right.
\end{equation}
where $x_j$ and $y_j$ are the fast and slow variables
respectively, $j=\overline{1,N}$. In all simulations we use
$\mu=10^{-3}$, $\alpha=3.5$, $\varsigma_i\in[0.15,0.16]$ (a
uniform random distribution)  that provides chaotic spiking in an
isolated map; $\varepsilon$ is the coupling strength, $K_i$ is the
number of entries in the $i$-th neuron. The sum is taken over all
neighbours of a neuron in the network; all connections are
reciprocal. The coupling function corresponds either to electrical
$G_{j,i}^k=x_{j}^k-x_i^k,$ or synaptical excitatory coupling
$G_{j,i}^k=(x_{rp}-x_i)\chi(x_j),$ here the reversal potential
$x_{rp}=1$, $\chi(x)=1$ if $x>0$, and $\chi(x)=0$ otherwise.

To analyze the collective dynamics of this neural ensemble in terms of
CPS one has to introduce frequency and phase characteristics of
oscillations. For spiking dynamics we determine the average spiking
frequency in neuron $j$ by:
\begin{equation}\label{eq3}
\omega_j = \lim_{k \rightarrow \infty} n_j^k/k,
\end{equation}
where $n_j^k$ is the number of spikes fired from the beginning up
to the discrete time $k$. The phase of spiking reads:
\begin{equation}
\label{eq4}
\begin{array}{l}
 \varphi_j^k
=2\pi\frac{k-k_{j,m}}{k_{j,m+1}-k_{j,m}}+2\pi m_j, k_{j,m} \le k
<k_{j,m+1},
\end{array}
\end{equation}
$k_{j,m}$ being the moment of the $m$-th spike in neuron $j$.

\begin{figure}[t]
{\centering
\resizebox*{0.90\columnwidth}{!}{\includegraphics{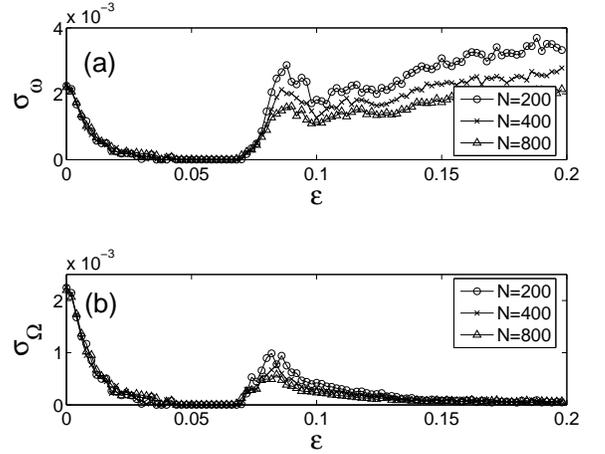}}}
{\caption{Variances of slow time-scale (STS) and spiking
frequencies $\Omega_j$ and $\omega_j$ over the chain (\ref{eq1})
vs. coupling strength $\varepsilon$ for different chain sizes
$N$.} \label{fig0}}
\end{figure}

Further on, we need the other characteristics, which correctly
describe the slow time-scale (STS) oscillations in the bursting
regime. The STS frequencies $\Omega_j$ and phases $\Phi_j$ are
defined similar to their spiking time-scale counterparts
$\omega_j$ (\ref{eq3}) and $\varphi_j$ (\ref{eq4}), except that
not each spiking event contributes to the $2\pi$ growth, but only
the first one in a burst (in simulations, the one coming after at
least $80$ iterations in silent state). Note, that while neurons
generate STS chaotic spiking (like for $\varepsilon=0$), both
definitions are equivalent. If fast repetitive spikes form trains
of bursts, $\Omega_j$ will characterize the bursting frequency and
$\omega_j$ will characterize the average spiking frequency. This
technique allows for a correct separation of the FTS and STS.

Now we summarize the regimes that occur for different coupling
strengths $\varepsilon$ in a regular {\it chain of electrically}
coupled neurons \cite{we_pre}. At low coupling neuronal firings
are unsynchronized (Fig.\ref{fig1}(a)), at moderate coupling they
get synchronized (Fig.\ref{fig1}(b)). As we increase
$\varepsilon$, the CPS regime becomes unstable and neurons start
firing fast repetitive spikes occasionally (Fig.\ref{fig1}(c)).
Further increase of $\varepsilon$ results in synchronized bursts
with a fractal-like spatio-temporal structure of spikes
(Fig.\ref{fig1}(d)) \cite{excitable}.

To quantify these transitions we have computed the variances of
the STS and spiking time-scale oscillations frequencies vs. the
strength of the electrical coupling for the chain lengths $N=200,\
400,\ 800$ (Fig.\ref{fig0}). We find three size-independent
critical coupling strengths: $\varepsilon_1\approx0.035$,
$\varepsilon_2\approx0.07$, and $\varepsilon\approx0.15$, which
define four intervals: (i) for $\varepsilon\in[0,\varepsilon_1]$
oscillations on the single existing time-scale -- the slow one --
are unsynchronized, (ii) for
$\varepsilon\in[\varepsilon_1,\varepsilon_2]$ oscillations on the
STS are synchronized, (iii) for
$\varepsilon\in[\varepsilon_2,\varepsilon_3]$ oscillations on the
second time-scale -- the fast one -- are initiated, both
time-scales demonstrate unsynchronized oscillations, (iv) and for
large coupling $\varepsilon>\varepsilon_3$ oscillations on the STS
get synchronized again, oscillations on the FTS are still
unsynchronized.

\begin{figure}[t]
{\centering
\resizebox*{0.90\columnwidth}{!}{\includegraphics{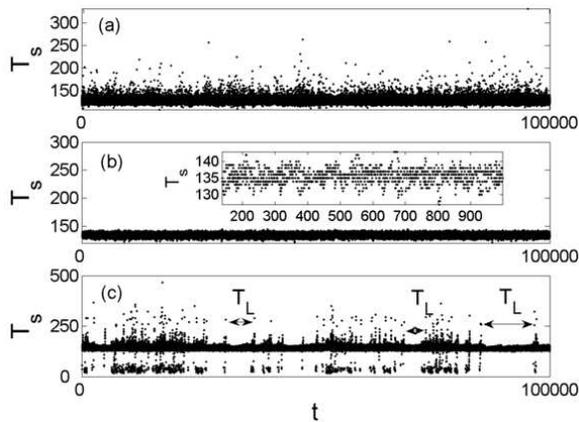}}
}
{\caption{Evolution of interspike intervals for (a)
$\varepsilon=0.01$, (b) $\varepsilon=0.05$, (c)
$\varepsilon=0.073$
In (c) we denote several intervals of stability windows $T_L$.}
\label{fig2}}
\end{figure}

\begin{figure}[t]
{\centering
\resizebox*{0.90\columnwidth}{!}{\includegraphics{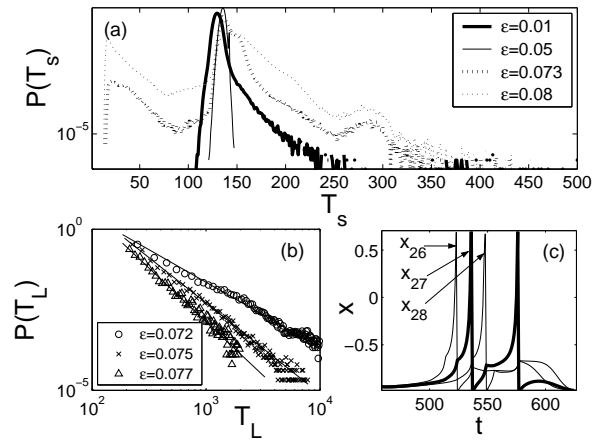}}}
{\caption{Probability distributions of (a) interspike intervals
$P(T_s)$ and (b) duration of stable CPS windows $P(T_L)$ for
different values of the coupling strength (solid lines are guides
for the eye). (c) Dynamics of fast variables $x_j$ in three
neighboured neurons. It is shown, how a spike in the $28$-th
neuron makes the $27$-th neuron (bold line) fire a repetitive
spike.}\label{fig3}}
\end{figure}

The transition to CPS at $\varepsilon_1$ is what one would have
intuitively expected {\it \'{a} priori}, as long as this is a
conventional way how arrays of non-identical oscillators behave
\cite{Pik_book}, but the instability of CPS at $\varepsilon_2$ and
the generation of the FTS by repetitive spikes further on demand a
detailed study. To uncover the nature of this transition, we
record the interspike intervals $T_s$ in each neuron and plot
their evolution for different coupling strengths (Fig.\ref{fig2}).
For $\varepsilon<\varepsilon_2$ we observe that chaotic spikes
construct only the STS ($T_s>100$), be it unsynchronized
(Fig.\ref{fig2}(a)) or synchronized (Fig.\ref{fig2}(b)) dynamics.
Note that in the synchronization regime the relative spike timing
in neurons is locked but not tightly fixed. It varies from one
spiking front to another one exhibiting flexibility of phase,
typical of genuine CPS (see the inset in Fig.\ref{fig2}(b)). $T_s$
sequences (Fig.\ref{fig2}(c)) demonstrate the intermittent nature
of the developing instability. The time intervals, during which
fast repetitive spikes are generated, are interrupted by windows
of synchronized STS spiking (in Fig.\ref{fig2}(c) one of them
lasts as long as $T_L\approx10,000$ iterations, which is of the
order of $100$ STS interspike intervals). The closer $\varepsilon$
is to $\varepsilon_2$, the larger become stable CPS windows $T_L$.
In Fig.\ref{fig3}(a),(b) we show statistical properties of
interspike intervals $T_s$ and time durations of stable windows
$T_L$, respectively. Remarkably, the probability distributions of
$T_L$ demonstrate a power-law dependence over five decades in a
{\it finite} interval of the coupling strength with
$\varepsilon$-dependent exponents.

The shown complexity arises on the micro-scale, when a spike in an
adjacent neuron makes a just fired neuron fire again
(Fig.\ref{fig3}(c)) should the coupling be strong enough. It is
important to underline the principal role of the individual
dynamics of oscillators in synchronization and desynchronization
processes. In the classical case of coupled quasiharmonic
oscillators the variables change smoothly and the coupling tries
to synchronize the systems all the time. In case of spiking
neurons, the coupling synchronizes them until the faster neuron
fires. Its firing is also a synchronizing event, as it pushes the
slower oscillator up. On the opposite, firing of the slower neuron
desynchronizes them, as it pushes the faster one up towards the
next firing. Varying $\varepsilon$ we change the balance between
synchronization and desynchronization and observe the instability
of synchronization when short desynchronizing intervals prevail.

In addition, we implemented {\it synaptical excitatory} coupling
instead of the {\it electrical} one and found no qualitative
difference with respect to the shown effects. We would like to
stress that the observed instability of the synchronous regime is
not an artefact of a time-discrete system. We have observed it in
ensembles of non-identical Hindmarsh-Rose neuronal oscillators too
\cite{extended}.

\begin{figure}[t]
{\centering
\resizebox*{0.90\columnwidth}{!}{\includegraphics{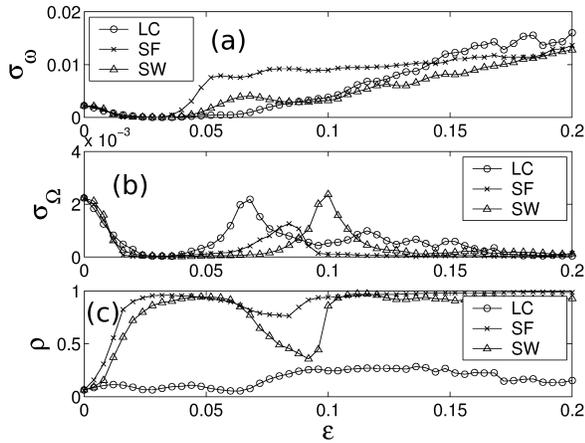}}}
{\caption{Variances of (a) STS and (b) spiking frequencies
$\Omega_j$ and $\omega_j$ and (c) the order parameter of the STS
oscillations $\rho$ vs. excitatory coupling strength for local
coupling (LC), scale-free (SF), and small-world (SW) network
topologies. Here $N=200$ and all data are averaged over $100$
realizations of network topologies and random parameters of
individual neurons.} \label{fig7}}
\end{figure}

Next we study whether these results are valid in complex neuronal
networks with a long-range synaptic connectivity \cite{note}, that
is typical of biological ensembles. In the following we implement
two types of complex topologies: scale-free and small-world ones
\cite{reviews}. Our interest in such networks has been
additionally stimulated by the recent study \cite{sf_experiment},
that reported scale-free properties of functional brain networks
(with the exponent varying from $2.0$ to $2.2$). The scale-free
network, we simulate, is characterized by the node degree
distribution $P(K)\propto K^{-\gamma}, \gamma=2.2$, and the mean
$<K>\approx4.2$. The small-world network has on the average $10$
links per neuron and the probability of rewiring a short-range
regular link is $p=0.1$. The results of the simulations
(Fig.\ref{fig7}(a,b)) demonstrate the same scenario of the onset
of bursting via instability of synchronized chaotic spiking. This
similarity becomes quite natural, as one takes into account, that
fast repetitive spiking is the result of interaction of two
neighbours, which fire with a mismatch in time, as discussed
above. Thus, the neighbour-to-neighbour interactions, and not the
global architecture, are important.

What for topology does matter, is the global coherence. Having in
mind, that precise timing of synchronized oscillations in
biological ensembles is considered to be a functional means in
cognitive tasks \cite{neur_review}, we calculate the order
parameter (as a measure of coherence) for the STS oscillations:
$\rho=|\sum e^{i \Phi_j}|$, Fig.\ref{fig7}(c). In complex networks
with long-range connectivity STS firing is tightly locked within
$10$ time durations of a single spike, while in locally coupled
ensembles the global coherence is absent
(Fig.\ref{fig7}(c),\ref{fig1}(b)). Thus the long-range coupling
strength, subjected to synaptic plasticity, appears to be a
plausible way for dynamical altering between coherent and
non-coherent performance, suggesting, in turn, a mechanism for
information processing in biological networks.

In summary, we have shown that ensembles of non-identical neurons
generate an instability of synchronous chaotic spiking, as the
coupling strength is increased. Arising spatio-temporal
intermittency gives birth to bursting dynamics, which at stronger
coupling becomes synchronous. This phenomenon has proved to be
generic with respect to the type of coupling and network
architecture. Beside general interest from the viewpoint of
theoretical nonlinear dynamics, these findings may directly apply
to neurobiological systems, indicating, that (i) excessive
coupling does not necessarily improve synchrony of spiking, and
(ii) the population dynamics can serve a mechanism behind
bursting, complementary to variation of individual parameters and
ionic mechanisms \cite{ionic}. And we strongly expect the reported
effects to be observed in biological experiments.

The authors thank M.I.Rabinovich for helpful discussions. M.I.
acknowledges financial support of INTAS YS Ref. Nr. 04-83-2816; G.O.
acknowledges RFBR 05-02-90567 and Gastproffessor des
Interdisziplin\"{a}ren Zentrum f\"{u}r Kognitive Studien f\"{u}r
Komplexer Systeme; and J.K. that of the International
Promotionskolleg Cognitive Neuroscience and BIOSIM.


\begin{thebibliography}{99}
\bibitem{extended} An extended paper comprising these results will follow.
\bibitem{note} Note, that electrical coupling between biological neurons
emloys gap junctions, i.e. elecrical connections are neccessarilly
short range ones, if not local.
\end{thebibliography}
\end{document}